# Evidence of gravitons as fused photons in four dimensions

Z R Adam
University of Washington Astrobiology Program
Box 352192, University of Washington, Seattle, WA 98195-2192

Email: voodrbar@u.washington.edu

June 1, 2009 (updated from original manuscript dated January 15, 2009)

**ABSTRACT**
A model of graviton momentum transfer was constructed to investigate a conjecture that gravitons are fused photons propagating in four dimensions. The model describes gravitational attraction between two bodies, each of simplified geometric shape and comprised of a calculable number of massive particles (quarks and leptons), as a probabilistic quantized mechanism of graviton scattering that gives rise to gravitational momentum flux. Earth-Human, Moon-Human, and Earth-Moon gravitational systems were investigated to solve for the wavelength of photons that comprise the graviton. The calculated wavelength for each system was approximately equal to the predicted value of the Planck length, which is interpreted as evidence that gravitons may be formed as fused four dimensional photons. The results corroborate current thinking about the temperature at which gravity separated from a unified force during the Big Bang, while explaining the weakness of the gravitational force from the atomic to the sub-planetary scale. Extension of the model produces unique, testable predictions arising from the averaged quantum properties of the graviton as fused photons, and the general model approach may be compatible with other efforts to describe the inner structure of the graviton.

**PACS Classification Categories:** 11.15.Kc, 11.25.Mj, 12.60.-i, 14.70.-e, 14.70.Pw, 14.80.-j.

## 1. INTRODUCTION

The graviton remains the most elusive of gauge bosons predicted by the Standard Model. The ability of the Standard Model to predict properties of elementary particles prior to their observation, and to explain perplexing observations made by particle physicists over the last decades, has repeatedly confirmed its utility for comprehending the basic structure of matter. The graviton, predicted to be a spin-2 massless boson based on properties of the metric tensor *[1]*, has thus far not been detected or measured with any certainty using the particle physics instruments at hand.

Particle theorists have speculated that foreseeable instruments cannot generate the energy densities necessary to create phenomena where a graviton signal can be created or observed *[1,2]*. Graviton thought experiments invoke Jupiter-sized detectors, with 100% detection efficiency, to yield results within a reasonable time of observation, since the mean free path of gravitons are presumably quite large *[3]*. But gravity is a unique problem in that successful field approximations have been derived prior to comprehending what may be a quantized phenomenon. In a sense, every massive body in the universe is a detection device with 100% efficiency. The universe itself is a living laboratory that can be used to compare our theoretical conjectures to observable astronomical phenomena.

The model in this paper builds entirely upon the conjecture that a graviton is a spin-2 massless boson that can be formed from two fused spin-1 photons, similar to that proposed as a Kaluza-Klein fusion process *[3]* but constrained to formation and propagation in only four dimensions. There is no loss of gravitons to higher dimensions or other manifolds- (3+1)D geometry produces results consistent with observed behavior. The simple energy-momentum exchange model described in this paper is naïve regarding the internal structure of gravitons (i.e., there is no quantum mathematical model for why photons have fused). Therefore the model results may be compatible with string theory, quantum field theory, loop quantum gravity, or other quantum frameworks that attempt to describe the internal structure of such a graviton.

To construct the model, assume that:

1) each graviton is comprised of two fused photons,

2) the energy and momentum of each graviton is the sum of the energy and momentum, respectively, of its constituent photons,

3) gravitons have no rest mass and may scatter other gravitons,

4) gravitons are 'pull-momentum' (i.e., always-attractive) particles,

5) the average net number of gravitons scattered by a massive body is approximately equal to the number of massive particles (quarks and leptons) that comprise that body

6) each massive particle is considered to have an averaged graviton cross-section, regardless of its mass,

7) massive objects may be considered macroscopically static with respect to one another.

The model has sufficient definition to test predictions concerning the nature of the conjectured graviton, and to make testable predictions about the graviton's observable effect on the structure of the universe.

## 2. METHODS

The graviton scattering model described in this paper is a billiards ball model of graviton momentum transfer. It functions by relating two properties of a di-photonic graviton to yield an estimate of the average wavelength of the photons that comprise the graviton. Gravitons must impart momentum between massive particles, and this momentum is related to the wavelength, $\lambda$, of its constituent photons through the quantized photonic energy equation $p=(2h/\lambda)$, with the factor of two arising because there are two photons that comprise the graviton. Newtonian physics provides a calculable rate of momentum transfer for a static body placed within a gravitational field that produces an acceleration on the object a, thus the number of gravitons imparting momentum can be related to the total momentum required, leaving $\lambda$ as a variable to yield the equation:

$$\left[\frac{dN}{dt}\right]_{GravityGradient} = \frac{m_2 a}{p_{NET}} = \frac{m_2 a}{\left[\frac{2h}{\lambda}\right]} = \frac{m_2 a \lambda}{2h}$$

On the other hand, the number of gravitons that a body will scatter is a probabilistic calculation based on a cross-section closely related to $\lambda^2$ and the total number of target areas associated with that body. The number of target areas within a body can be found by counting the number of massive particles within an object and assume an average graviton scattering cross-section per particle that scales with $\lambda^2$ (which is also left as a variable). The frequency rate of graviton movement through a stationary surface, perpendicular to a massive object creating graviton flux, is a function of the density of massive particles within the object and the speed of the gravitons (c, the speed of light) they are scattering:

$$\left[\frac{dN}{dtdA}\right] = n\frac{c}{4}$$

The probability, P, of Object 2 stopping a graviton may be estimated with the product of the massive particle density of Object 2, the graviton cross-section, and a characteristic absorption length of the Object 2:

$$P = n_2 \sigma_{Graviton} dx_2$$

The probability of scattering a graviton is then a product of the probability of scattering the graviton flux from Object 1, the area of flux through Object 2 through the Beer-Lambert Law, and the frequency rate of graviton movement through the area of flux attributable to Object 1:

$$\left[\frac{dN}{dt}\right]_{Scatter} = P\left[\frac{dN}{dtdA}\right]_1 dA_2 = (n_2 * A\lambda^2 * dx_2)\left(n_1 \frac{c}{4}\right) dA_2$$

A is a scalar relating the interaction cross-section of the graviton to the square of the wavelength of light that comprises it: $\sigma_{Graviton} = A\lambda^2$. An exact value of A=4.6106 can found by equating the Stefan-Boltzmann equation to the number of photons in a box, reducing the number of photons to 1, substituting Wien's Displacement Law for the product $\lambda T$ to solve for a characteristic volume, and then adding two of these characteristic volumes together (as a function of wavelength) into a sphere whose cross-sectional area is then calculated to yield A. This solution is consistent with Heisenberg's Uncertainty Principle because the exact location of the photon within this volume is unknown, and the photon's energy function may be assumed to be smeared out within this volume. This approach to finding the graviton cross-section is comparable to that taken by Bekenstein and Mukhanov [4] when examining the quantization of black hole surfaces. This scalar will be varied from 1 to 10 to establish a range of uncertainty, however, it is expected that the final calculated wavelength values computed using 4.6106 should fall very close to the predicted value. This is because there is a very close relationship between the temperature of a reaction and the most probable wavelength of light produced, and this is encapsulated within the use of Wien's Displacement Law to derive the characteristic volume.

By equating the number of gravitons required to impart momentum on a body in a gravitational field to the number of gravitons scattered by that body, the result is a system with the variable $\lambda$ as an estimate of the wavelength of photons that comprise the conjectured di-photonic graviton. Equating the two graviton rate equations, substituting for the acceleration according to Newton's equation ($a = Gm_1/r^2$) and solving for the wavelength, $\lambda$, yields:

$$\frac{m_2 m_1 G \lambda}{2hr^2} = (n_2 * A\lambda^2 * dx_2)\left(n_1 \frac{c}{4}\right) dA_2$$

$$\lambda = \frac{2 m_2 m_1 G}{n_2 n_1 A ch dx_2 dA_2 r^2}$$

The relationship between the mass of an object, $m_x$, and the number of massive particles (quarks and leptons) that comprise that object, $N_x$, is dependent on which atomic species comprise the object, as demonstrated in Table 1 below.

**Table 1. Tabulation of calculation of the linear factor B, relating the number of massive particles of a material to the mass of that material.**

| Massive Object | Number of Protons | Number of Neutrons | Number of Electrons | Number of Massive Particles | Atomic Mass [kg/mole] | Mass Per Unit [kg/unit] | B [massive particles/kg] |
|---|---|---|---|---|---|---|---|
| Electron | 0 | 0 | 1 | 1 | - | 9.11E-31 | 1.10E+30 |
| Proton | 1 | 0 | 0 | 3 | - | 1.66E-27 | 1.79E+27 |
| Neutron | 0 | 1 | 0 | 3 | - | 1.66E-27 | 1.79E+27 |
| H | 1 | 0 | 1 | 4 | 0.001 | 1.66E-27 | 2.41E+27 |
| He | 2 | 2 | 2 | 14 | 0.004 | 6.64E-27 | 2.11E+27 |
| Li | 3 | 4 | 3 | 24 | 0.007 | 1.16E-26 | 2.06E+27 |
| C | 6 | 6 | 6 | 42 | 0.012 | 1.99E-26 | 2.11E+27 |
| N | 7 | 7 | 7 | 49 | 0.014 | 2.32E-26 | 2.11E+27 |
| O | 8 | 8 | 8 | 56 | 0.015999 | 2.66E-26 | 2.11E+27 |
| Na | 11 | 12 | 11 | 80 | 0.02299 | 3.82E-26 | 2.10E+27 |
| Mg | 12 | 12 | 12 | 84 | 0.024305 | 4.04E-26 | 2.08E+27 |
| Al | 13 | 14 | 13 | 94 | 0.026982 | 4.48E-26 | 2.10E+27 |
| Si | 14 | 14 | 14 | 98 | 0.028086 | 4.66E-26 | 2.10E+27 |
| S | 16 | 16 | 16 | 112 | 0.032065 | 5.32E-26 | 2.10E+27 |
| Ca | 20 | 20 | 20 | 140 | 0.040078 | 6.66E-26 | 2.10E+27 |
| Fe | 26 | 30 | 26 | 194 | 5.58E-02 | 9.27E-26 | 2.09E+27 |
| Ni | 28 | 31 | 28 | 205 | 0.058693 | 9.75E-26 | 2.10E+27 |

For most charge-neutral objects such as the Earth, the Moon, and human beings, the relationship can be simplified to $m_x = \frac{N_x}{B}$, where B is approximately equal to 2.09E+27 massive particles per kg material based on elemental assays of these objects. This corresponds to the simplified case for most stable atomic species that comprise rocky material, where the ratio of protons:neutrons:electrons is 1:1:1.

$$B_{Generic} = \left[ \frac{N_{proton} m_{proton} + N_{neutron} m_{neutron} + N_{electron} m_{electron}}{N_{quarks} + N_{leptons}} \right]^{-1} \approx 2.09E+27 \left[ \frac{massive\_particles}{kg} \right]$$

Substituting this linear relationship in place of the masses of Objects 1 and 2 yields the general equation for of the di-photon graviton wavelength calculable for two massive bodies comprised of a calculable number of massive particles:

$$\lambda = \frac{2G}{n_2 n_1 A c h d x_2 d A_2 r^2} \frac{N_1 N_2}{B^2} = \frac{2G}{n_2 n_1 A c h d x_2 d A_2 r^2} \frac{n_1 V_1 n_2 V_2}{B^2} = \frac{2 G V_1 V_2}{A B^2 c h d x_2 d A_2 r^2}$$

This leaves five variables to solve for this generic system: the volume of Object 1 ($V_1$), the volume of Object 2 ($V_2$), the characteristic absorption length of Object 2 ($dx_2$), the flux area of gravitons from Object 1 through Object 2 (dA$_2$), and the distance between the centres of Objects 1 and 2 (r).

This is the general form of the equation that will be used to evaluate two cases:

> Case 1: A cylindrical model of a human being standing on the surface of a large, spherical object (the Earth or the Moon).
>
> Case 2: A model of the Moon orbiting around the Earth.

*2.1. Case 1: Modeling Earth-Human and Moon-Human Systems*
For the Earth-Human model sub-case, the Earth (Object 1) is a sphere of radius 6378.1 km (V$_1$ = 1.0868E+21 m$^3$). The human being (Object 2) is modeled as an upright cylinder, 70 kg in mass with a density of 1 g/cm$^3$, 11.3 cm in radius, 1.75 meters in height (V$_2$ = 0.070 m$^3$), comprised of a number of massive particles related to the constant B described above. The flux area of Object 2 is the addition of the flux of the lower circular base and the side of the cylinder (dA$_2$ = πR$_2^2$ + 2πR$_2$H$_2$ = 1.2885 m$^2$). The average absorption distance through Object 2, dx$_2$, is equal to about 1.052 m, but will be varied from 22.6 cm (2 x R$_2$) to 1.75 m to establish upper and lower bounds for the final wavelength result. The distance between objects, r, is simply equal to the radius of the Earth (r = 6378.1 km).

The Moon-Human sub-case is modeled in exactly the same way as the Earth-Human sub-case, except that parameters for the Moon are used (R$_1$ = 1737.4 km, V$_1$ = 2.1968E+19 m$^3$). The distance between objects, r, is equal to the radius of the Moon (r = 1737.4 km).

*2.2. Case 2: Modeling an Earth-Moon Gravitational System*
For the Earth-Moon model, two sub-cases are considered: the background graviton density of space is either much lower or much higher than the graviton density of Objects 1 and 2. These possibilities have implications regarding the larger picture of the universe since no graviton background density has ever been detected. If the graviton background is low, then the measured physical extent of the Moon corresponds directly to the graviton scattering length and flux area in a similar manner described in the planet/human systems in Case 1. If the graviton background density is high, then the Moon's apparent graviton interaction properties (absorption length and flux area) may be much larger than its observable, physical dimensions of radius and exposed surface area. A body with a larger graviton flux in the vicinity of the Moon (i.e., the Earth and the Sun) would cause the flux of gravitons from the Moon to reach equilibrium with its surrounding environment. At a calculated 'apparent' radius, the flux of gravitons toward the Moon from the Earth would equal the flux outward from the Moon itself, reaching graviton flux equilibrium, as roughly depicted in Figure 1 below. This is akin to the equilibrium reached between the charged particle flux from the Sun and Earth's magnetic field, which extends far beyond the physical extent of the planet itself and prevents solar particles from impinging on the atmosphere.

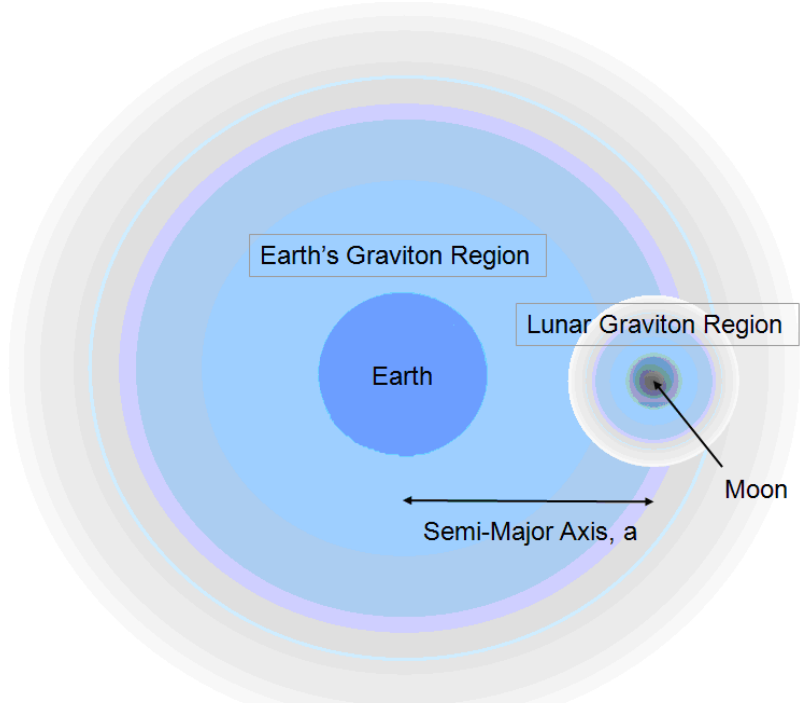

**Figure 1. Depiction of the graviton densities surrounding the Earth and the Moon (not to scale).**

For the sub-case of low graviton background density of space, Earth is modeled as a sphere of radius 6378.1 km ($V_1 = 1.0868E+21$ m³). The Moon is modeled as a sphere of radius 1737.4 km ($V_2 = 2.1968E+19$ m³). The flux area of Object 2 is the projected area facing Object 1 ($dA_2 = \pi R_2^2 = 9.483E+12$ m²). The average absorption distance of Object 2, $dx_2$, is equal to $(4/3)R_2 = 2316.5$ km, calculated by inscribing the Moon into a cylinder with an equal radius and equal volume to that of the spherical Moon. The distance between objects, r, is simply equal to the average distance between the Earth and the Moon ($r = 3.844E+8$ m), which will be varied by +/- 5% to account for the variation in centre-to-centre distance along the Moon's orbit, to produce a range of uncertainty for this model.

For the sub-case of a high graviton background density in space, Earth is again modeled as a sphere of radius 6378.1 km ($V_1 = 1.0868E+21$ m³). The 'apparent radius' of the Moon is expanded to be the radius at which the graviton flux from the Moon is equal to the graviton flux from the Earth at the Moon's average distance, which can be calculated (in terms of total numbers of gravitons, $N_x$, or in terms of graviton densities, $n_x$, of Objects 1 and 2) as:

$$\left(\frac{N_2}{V_2}\right)\left(\frac{c}{4}\right)\left(\frac{4\pi R_2^2}{4\pi R_{app}^2}\right) = \left(\frac{N_2}{V_2}\right)\left(\frac{c}{4}\right)\left(\frac{4\pi R_1^2}{4\pi R^2}\right)$$

$$R_{app} = R\left(\frac{N_2 R_1}{N_1 R_2}\right)^{\frac{1}{2}} = R\left(\frac{R_2}{R_1}\right)\left(\frac{n_2}{n_1}\right)^{\frac{1}{2}}$$

For the Earth/Moon system, the apparent radius $R_{app}$ is 81580 km, which is slightly larger than the Hill Sphere of the Moon. The flux area of Object 2 becomes the apparent projected area facing

Object 1 ($dA_2 = \pi R_{app}^2 = 2.09E+16$ m$^2$). The average absorption distance of Object 2, $dx_2$, is equal to $(4/3)R_{app} = 108774$ km, calculated by inscribing a sphere with radius $R_{app}$ into a cylinder with an equal radius and equal volume. The distance between objects, r, remains equal to the average distance between the Earth and the Moon (r = 3.844E+8 m), which will also be varied by +/- 5% to account for the variation in centre-to-centre distance along the Moon's orbit, to produce a range of uncertainty for this model. However, a new term must be added to account for the number of background gravitons scattered by the Moon within this larger sphere. It takes light $t_{app} = R_{app}/c$ seconds (about 0.272 s) to travel the distance from the Moon to the edge of the apparent sphere. In this time, the number of scattered gravitons is:

$$N_{2,Emitted} = \left(\frac{N_2}{V_2}\right)\left(\frac{c}{4}\right) t_{app} (4\pi R_2^2) = N_2 \frac{3R_{app}}{4R_2}$$

These values can then be substituted into the general equation for the wavelength of light that comprises the graviton.

### 3. PREDICTIONS

There are a number of heuristic thought experiments that predict that electromagnetic and gravitational phenomena unify as the wavelength of a photon approaches the Planck length *[5,6]*, since at this wavelength a quantum of light reaches an energy density comparable to a black hole of about this size. Its wavelength is roughly equivalent to its Schwarzschild Radius (or more appropriately, the radius of a Michelle-Laplace dark body) *[7]*. The energy density of this photon is comparable to the energy density at which gravity became a distinct force following the Big Bang *[5,6]*. Equations that yield the Planck length as a characteristic wavelength are based on heuristic thought experiments (or dimensional analysis) and very little more *[7]*. These thought experiments illustrate the limits of our current understanding of physical processes, but currently they do not provide any description of a mechanism relating quantum and gravitational phenomena, as they do not describe how gravity is produced.

It is predicted:

1) The wavelengths for the Earth-Human and Moon-Human systems will be approximately equal to the Planck length (1.616E-35 m or 4.051E-35 m, depending on whether one uses h or ℏ for the dimensional analysis equation) *[7]*, as is consistent with Big Bang theory and the separation of gravity from a unified force.

2) The wavelength for the Earth-Moon system will be approximately equal to the Planck length for only the high background graviton density sub-case. The wavelength for the low background density sub-case will deviate substantially from the Planck length because this description would not fit the observed smoothness of the gravity gradient, and gravitational retention of gaseous species near massive bodies would require additional explanation in the context of this sub-case.

## 4. RESULTS

All calculated results for the wavelength of light that comprise di-photonic gravitons were within the predicted range near the Planck length. The Earth-Moon system analyzed assuming a high graviton background density yields a value of approximately the Planck length, while the low graviton background density system yields a result that deviates from the Planck length by a factor of about 100. The results are presented in Figure 2.

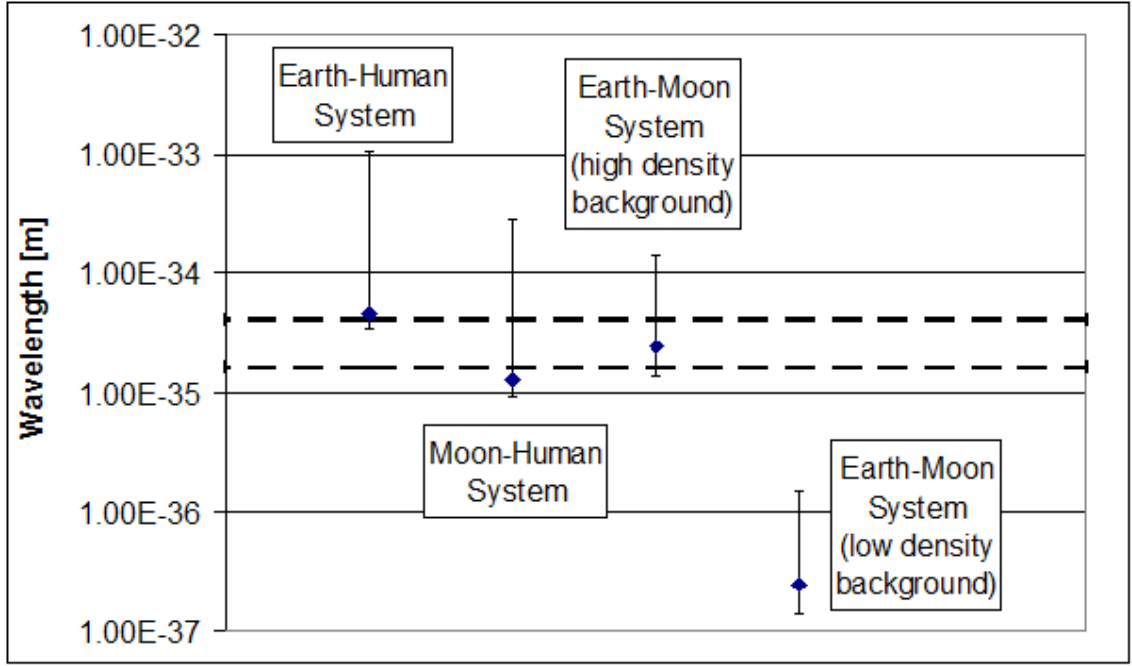

**Figure 2. Depiction of the results of each system analyzed. The heavy bars denote the Planck length as calculated with h and ℏ.**

## 5. DISCUSSION

The wavelengths of the photons comprising gravitons corroborate conjectures made by Big Bang physicists, roughly matching the timeline of separation of gravity from a unified force about 1 Planck second after the Big Bang. No phenomena have ever been created in a lab at these energy densities. The fact that a crudely-built graviton model can corroborate predictions near the Planck length may be significant, if only to illuminate the reason for the apparent weakness of gravity compared to the other fundamental forces. Gravitons, as individual quanta of energy, are actually far more energetic than all other gauge bosons. But the tiny cross-section of interaction, combined with an apparent high universal background density, means that the force appears weaker when observed from the atomic to the sub-planetary scale of observation.

A review paper discussing ill-founded conceptions about the Planck length *[7]* points out that, to date, there are no physical models or observational evidence implying the existence of 'Planck particles'. The calculated results of this model provide at least one counterargument. Unlike the dimensional analysis and flawed equivalences (e.g., equation of Euclidean and non-Euclidean quantities) described in the review paper, the approach described in this paper derives the Planck length as being fundamental to a mechanism that explains gravitational field equations at the macroscopic scale, resulting in a simple system of equations with one unsolved variable: the wavelength of the photons that comprise a graviton. There are no automatic reasons why the wavelength should converge at the Planck length, it is simply the only wavelength that fits the observed data. This model assumes that a quantized mechanism is at work, and the calculated results match the concepts described by the heuristic models.

The success of the high background graviton density sub-case in which the Earth-Moon system wavelength is returned to the Planck length may imply that the background graviton density of the universe is much higher than the massive particle density of objects such as the Earth and Moon. The model incorporates a type of 'standing bubble' approach, which requires finding the radius at which the flux outward from the Moon equals the flux inward from the Earth, and any graviton interactions within that bubble contribute to the Earth's pull on the Moon. A high density background would explain this succinctly, as this background would propagate the net effects of these graviton-graviton interactions back to the source body much the same way that the Earth's crust propagates surface impacts as sound waves throughout its entire body. If there exists a high density, isotropic graviton background, then gravitons are neither created nor destroyed by masses such as the Earth, Moon, human beings, etc. These objects scatter the pre-existing isotropic background. By scattering the background, information about the location, physical dimensions, and mass of the object is propagated away from the body through graviton-graviton collisions.

The isotropic background smoothes the effects of gravity at all scales larger than the mean free path (MFP) of the graviton background. This would imply that the graviton MFP must be significantly shorter than the Bohr radius. Otherwise, one is forced to explain how gaseous atoms and molecules confined to massive bodies by gravitational forces scatter gravitons on unreasonably long lifetimes *[8]*, and how graviton interactions with individual electrons do not rip them out of stable atomic orbits. One possible reason why an isotropic graviton background has not been measured is because such 'Planckian' black holes would have a Hawking Temperature of 0 K, indicating that entropic effects that would cause thermal energy transfer properties to be observed and measured between gravitons and normal matter may not apply *[9]*. This would imply that the background graviton density constitutes a contribution to the zero point energy of the universe.

If the calculated results are no coincidence, then the implications of the model are avenues for testing and extending this theory of quantum gravity. No other models of quantum gravity predict the existence of so many isotropic background gravitons. There may be multiple mechanisms that produce graviton flux gradients. If one accepts that a massive object constantly scatters background gravitons that leave the object at the speed of light, then as the object approaches the speed of light a graviton 'shock wave' must form in front of the object in the direction of forward travel. This graviton gradient build-up would cause light approaching the object from the direction of travel to be blue-shifted, just as is predicted by the Theory of Relativity. A related implication is that a moving object must emit more gravitons than an identical object in a rest frame of reference (since it will scatter higher numbers of the isotropic graviton background), thus perhaps providing an explanation for the increase in the relativistic mass of an object that scales with velocity.

*5.1 Unique Predictions through Extension of the Model*
There are predictions that may be formulated through the use of this model that other quantum gravity models cannot produce. These predictions illustrate that the proposed mechanism is consistent (internally and with respect to other physical mechanisms) and may offer unique means of advancing our knowledge of gravitons.

*5.1.1 Minimum Possible Radii of Gravitationally-dominated Rocky Bodies.* There should be an observable effect on macroscopic bodies if there is a Planck-scale cutoff in permissible photon wavelengths that comprise a graviton. Percacci *[10]* has discussed the implications of finding evidence for such a cutoff. Extension of this model predicts that there should be a corresponding cutoff in the minimum size of macroscopic bodies that are formed under the dominating influence of gravity that can be exactly attributed to the properties of the graviton.

A brief summary of the model extension is as follows. Assume that Object 1 is a massive, static, charge-neutral sphere of radius $R_1$, comprised of $N_1$ massive particles (quarks and leptons), and that a smaller Object 2 is a sphere of radius $R_2$ with $N_2$ massive particles under the gravitational influence of Object 1. It was demonstrated that there is a simplified linear relationship between the mass of an object, $m_x$, and the number of massive particles that comprise that object, $N_x$, that goes as $N_x = Bm_x$, where B is equal to 2.09E+27 gravitons per kg for massive materials comprised mostly of elements heavier than hydrogen.

The acceleration due to gravity experienced by Object 2 due to Object 1 is in meters per second squared, which can be related to the masses of Objects 1 and 2, the distance between their centres, and their total respective numbers of massive particles through the relationship with the contstant B above and Newton's Law of Gravitation. Gravitons are modeled as spherical objects with cross section $A\lambda^2$ (where A was found to be approximately equal to 4.6106 by comparison with Wien's Displacement Law for an individual photon). The momentum of each graviton-graviton interaction is therefore $p=(2h/\lambda)$ from the quantized photon energy equation. The net number of gravitons fluxing through the surface of Object 1 is equal to $(N_1/V_1)*(c/4)$, where c is the speed of light. An approximation is made that the product of the flux area and the characteristic absorption length of Object 2 is roughly equal to the volume of Object 2. Equating the number of graviton momentum transfer events that must occur per second to produce the gravity gradient experienced by Object 2 to the number of graviton scattering events that occur per second with the given flux of gravitons through Object 2 yields a quadratic equation with two possible solutions for the wavelength, $\lambda$, of the light that comprises the graviton; null, and the relationship:

$$\lambda = \frac{2GV_1}{AB^2 chr^2}$$

where r is the centre-to-centre distance between Object 1 and Object 2.

Considering the special case of a static, spherical Object 1, with Object 2 resting on its surface, $V_1$ is equal to $(4/3)\pi R_1^3$ and r is equal to $R_1$. This reduces the equation to:

$$\lambda = \frac{8\pi GR_1}{3AB^2 ch}$$

Rearranging for $R_1$ yields:

$$R_1 = \frac{3\lambda AB^2 ch}{8\pi G}$$

This equation indicates that the shorter the wavelength of the photons that comprise a graviton, the smaller the characteristic radius of a spherical, static, homogeneous solid body that is held together with gravity as a dominating force.

The assumptions about Object 1 used to build the model are:

1) Spherical shape
2) Possessing a homogeneous graviton (mass) distribution throughout its volume
3) Static (not rotating or moving with appreciable relative speed)
4) Not being acted upon by any other internal or external (i.e., tidal) forces

If a minimum possible (cutoff) wavelength for photons is imposed, then a minimum corresponding cutoff radius for a spherical, homogeneous, static, massive body may be predicted by this model, since the model is expected to break down for real bodies that deviate from these assumptions. This implies that if photonic wavelengths could be shorter than the Planck length, then gravitons would have more momentum, and thus a larger binding strength to hold smaller bodies together in a spheroid shape. The minimum wavelength assumption is supported by theoretical evidence that the Hawking temperature goes to zero for Planck mass black holes, implying stable black hole particles of this size would not evaporate. *[9]* Photons with wavelengths approaching the Planck length approach energy densities comparable to Planck mass black holes *[3]*, and therefore would be stable structures that would persist, undetected, throughout all space. The specific spheroid cutoff radius for a given mix of meteoritic materials (such as dust, ice, pebbles, and the like) will differ based on the electrostatic bonding strength of the materials that comprise Object 1, but the transition range itself should be dominated by the character of gravitons, not the materials that make up the object.

Assuming that the Planck wavelength is the shortest possible wavelength of light that can comprise a di-photonic graviton, it is predicted that the corresponding transition radius calculated with the extension of the graviton radiation model will match the transition radius for spheroidal to non-spheroidal massive bodies in our solar system. Substituting a value for the Planck length for the minimum possible wavelength (using 4.05E-35 m for the Planck length, which matches the results of the gravitational systems investigated above) yields a characteristic cutoff radius for spheroidal massive bodies of 290000 m. This cutoff range can be compared to confirmed data concerning the size and shape of rocky planets, dwarf planets, moons, and asteroids. It is expected that bodies larger than this cutoff range would be spheroidal in shape (excepting those under the influence of excessive rotational or tidal forces) and bodies smaller than this cutoff range are predicted to possess mostly non-spheroidal shapes, deviating further from a regular shape as the bodies become much smaller than the cutoff range.

Raw data was collected to compare the size and shape of planets, dwarf planets, moons, and asteroids. An object was classified as 'spheroidal' if the ratio of its minimum axis to its maximum axis was greater than 0.91. An object was classified as 'ellipsoidal' if the ratio of its minimum axis to its maximum axis fell within a range between 0.5-0.91 and its two smaller axes were approximately equal. An object was classified as 'irregular' if the ratio of its minimum axis to its maximum axis was less than 0.91, and its two smaller axes were not approximately equal. The table of data used to generate this plot is available upon request. The data is restricted to information about the size and shape of bodies that have been confirmed by pictures or lightcurve data. If the specific shape of the body was not known, then it was not included in the data table. The data table includes size and shape information for 111 rocky solar system objects (planets, dwarf planets, moons, and asteroids), ranging from the largest object with a radius of 6371 km (Earth) to the smallest object with a radius of about 502 m (3908 Nyx).

The observations of the cutoff for spheroid objects match the predicted cutoff range values very well. The results for the min/max ratio of axes for each body are plotted versus the mean radius of each body in Figure 3 below, along with a heavy dashed line that corresponds to the predicted cutoff radius based on the imposed UV cutoff for the graviton. Almost all bodies with radii higher than the maximum predicted cutoff radius are spheroidal. Almost all bodies with radii lower than the minimum predicted cutoff radius are irregular or ellipsoidal.

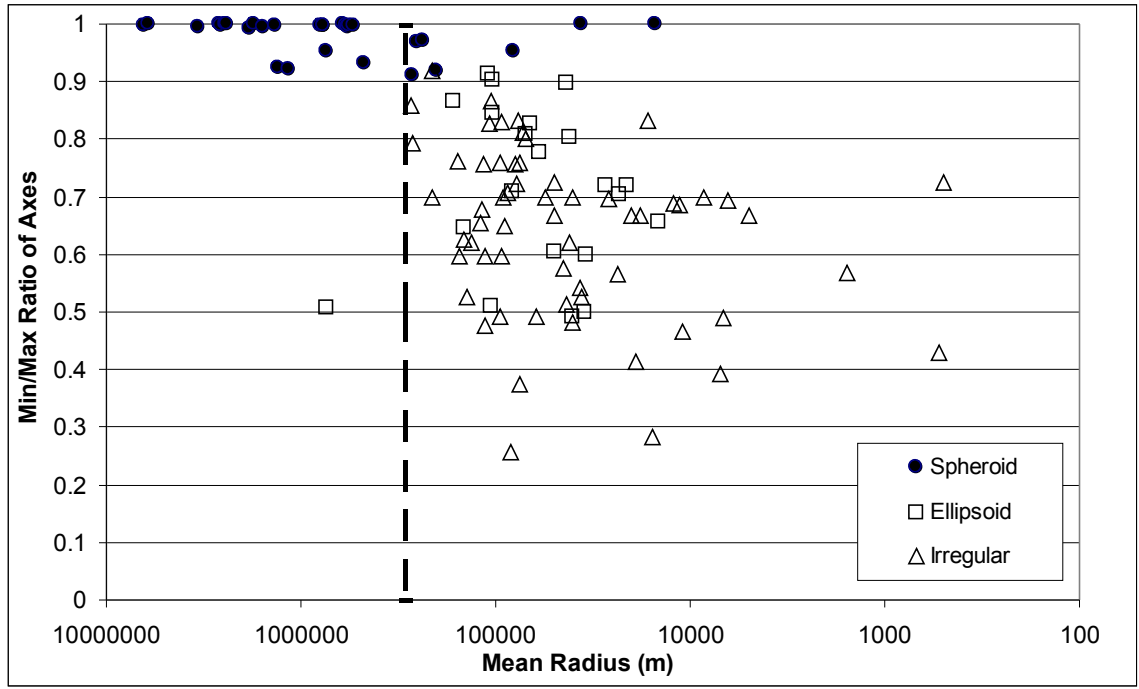

**Figure 3. A diagram depicting the ratio of minor to major axes versus mean radius for solar system rocky bodies. The heavy dashed line corresponds to the predicted cutoff radius for the transition from spheroidal to non-spheroidal bodies based on the properties of the graviton.**

There is one outlying rocky body that is non-spheroidal with a mean radius in excess of the predicted cutoff: the ellipsoidal body Haumea (r = 739000 m). However, this body is confirmed not to be rotationally static. Haumea has an exceptionally fast rotation period of about 3.9 hours, causing its extreme elongation such that its minimum/maximum axes ratio is about 0.51, even though the body is in hydrostatic equilibrium. This is a case wherein the rotation of the body causes dynamic deformation of the object, most likely due to a large impact with another asteroid.

There are two irregular bodies plotted near the upper calculated limit of 290000 m: the asteroids 2 Pallas and 4 Vesta both have moderately fast rotation periods as well (7.8 and 4.3 hours, respectively) which contribute to their elongation and deformation into a non-spheroidal shape. In addition, 4 Vesta has undergone numerous massive impacts which are theorized to have contributed to its deformation. Both bodies are dwarf planet candidates, if it can be shown that they have relaxed into hydrostatic equilibrium, but neither body may be considered completely static.

*5.1.2 Isotropic Graviton Background Density of the Universe.* A prediction arising from use of this model is that the isotropic background graviton density of the universe is high, with a graviton MFP significantly shorter than the Bohr radius in 'empty' space. This is predicted based on comparison of results of the Earth-Moon sub-cases, the stability of planetary atmospheres and electron-nuclei orbits, and because the gravity gradient appears to be very smooth. If the background were low in density, the calculated graviton interaction rate for everyday objects would be measurable as weight fluctuations. For example, objects with mass significantly less than about 1 kg would have graviton scattering rates less than 1 graviton per second on the Earth's surface, however these objects do not oscillate in measured weight with this frequency.

An extension of this model can be used to establish a calculable upper limit for the background graviton density of the universe. Consider a rocket-powered elevator with a combined structure and payload of mass $m_1$, and a propellant mass of $m_2$, which is initially at rest, as depicted in Figure 1a. The rocket is then fired to deliver an impulse burn of the shortest possible duration, as mass $m_2$ is expelled out of the thrust chamber at a velocity $v_2$ at an impulse that accelerates the propellant over distance $\Delta x$, as depicted in Figure 4. The shortest possible distance through which to accelerate the propellant to generate thrust is assumed to be the Planck length (4.05E-35 m). This is a limit that is consistent with evidence for a fixed UV cutoff for both photons and gravitons. There is no physical meaning for phenomena to propagate at distances less than this cutoff:

$$\Delta x > l_{Planck} = 4.05E-35 [m]$$

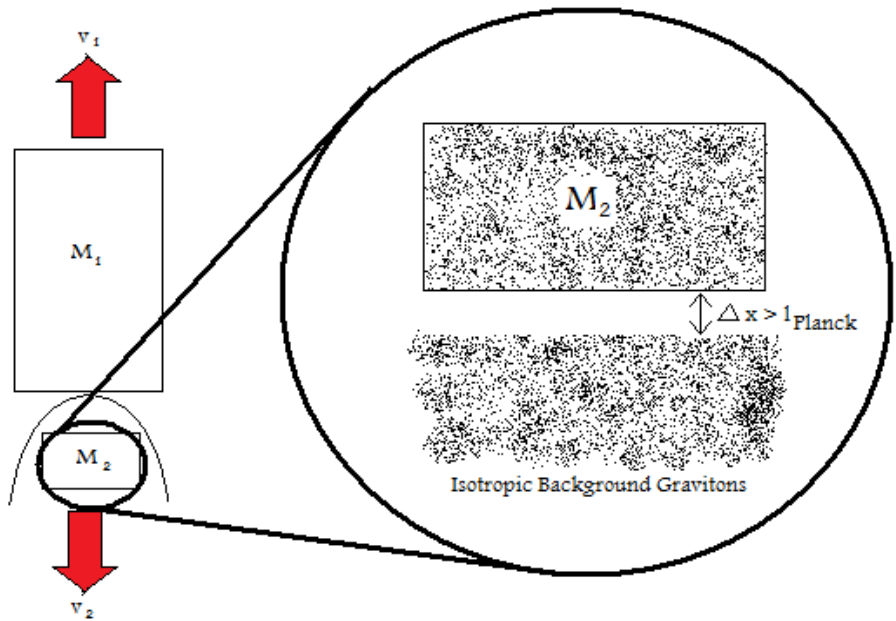

**Figure 4. A diagram of the rocket-powered elevator setup used to estimate the isotropic background graviton density.**

The payload and structure of mass $m_1$ is propelled in the opposite direction at $v_1$. A probability of interaction for each massive particle in the propellant mass may be calculated, with the density of the universal background of gravitons, $n_{Universe}$, left as an unknown:

$$P = n_{Universe} \sigma_{Graviton} dx$$

The number of particles that are expected to interact with the graviton background can be related to the probability of interaction multiplied by the total number of massive particles subjected to this probability:

$$dN = N_2 P = N_2 n_{Universe} \sigma_{Graviton} dx$$

On the other hand, conservation of momentum yields the number of gravitons that are required to be scattered to generate thrust. In this view, mass is a measure of the ability of a substance or quantity to generate a graviton gradient when accelerated. Therefore, for momentum to be conserved the following relationship must be true:

$$m_1 v_1 = m_2 v_2 = dN m_{Planck} c$$

This can be rearranged, using the linear relationship between mass and the number of graviton-scattering massive particles through the constant, B=2.09E+27 [massive particles/kg] to yield:

$$\left(\frac{N_2}{B}\right) v_2 = dN m_{Planck} c$$

$$dN = \frac{N_2 v_2}{B c m_{Planck}}$$

Equating the two equations describing the number of graviton interactions and solving for the background graviton density of the Universe yields:

$$n_{Universe} = \frac{1}{\sigma_{Graviton} B m_{Planck} dx}$$

$$dx > l_{Planck}$$

$$n_{Universe} < \frac{1}{B \sigma_{Graviton} m_{Planck} l_{Planck}}$$

The graviton cross section has been previously estimated to be $4.6106(\lambda_{Planck})^2$. Substituting for all known values yields:

$$n_{Universe} < 8.83E+82$$

A value for the mean free path (MFP) can now be calculated with all known values:

$$MFP = \frac{1}{\sigma_{Graviton} n_{Universe}}$$

$$MFP > 1.50E-15 m$$

This distance is approximately equal to the maximum range of the strong nuclear force, and is much smaller than the Bohr radius. This implies that deviations from a field description of gravity may be expected to break down at distances less than the range of the strong nuclear force.

*5.1.3 Gravity Waves as Sound Waves in an Isotropic Graviton Gas.* The concept of the isotropic graviton background as a 'perfect gas' can easily be subjected to a consistency check: would this gas propagate gravity waves at the speed of light, as implied by the Theory of Relativity? The answer is yes. In this case, gravity propagation is analogous to a sound wave in the gas (a propagation of a graviton gradient through graviton-graviton collisions). Gravitons have 2 degrees of freedom *[11-13]*, thus the ratio of specific heats (γ) is also 2. The equivalent rest mass of the propagating particles is twice the Planck mass (one for each photon that comprises the graviton), and the temperature is the Planck temperature (corresponding to the energy density of the graviton and the temperature at which it separated from a unified force). The test is passed: gradient disturbances propagate at the speed of light:

$$Speed_{Gravity} = \sqrt{\frac{\gamma T_{Planck} k_B}{2 m_{Planck}}} = c$$

*5.1.4 A Physical Basis for Quantum Uncertainty.* Another prediction ties quantum uncertainty to the graviton background. This paper has attempted to describe only what happens at the macroscopic scale due to graviton/massive particle interactions. One possibility is the prediction that quantum uncertainty is regulated, and perhaps even produced, by the constant interaction of isotropic background gravitons with individual atoms and molecules (called 'zitterbewegung' by Schrödinger) *[14]*. It is already commonly accepted that the universal 'vacuum' is no vacuum at all, but a 'boiling sea of quantum fluctuations…' capable of rapid, short-duration changes in energy content and particle number *[15]*. Interaction with a graviton background would impart discontinuous and unpredictable variation in speed and momentum that the Uncertainty Principle predicts for atomic particles, while the net force on the particle is zero over long time periods. These momentum jumps are sufficient to overcome confinement barriers at the atomic or subatomic scales. In this light, Heisenberg's Uncertainty Principle may be a description of Brownian motion of atomic and subatomic particles, with the same universal isotropic graviton background 'gas' that smoothes out the graviton gradient of large, massive bodies. The model described in this paper may provide a means of connecting the relationship between gravitational momentum flux, quantum entropy, and the square of the Planck length through the momentum and cross-sectional area of gravitons that propagate the gravitational force, providing a physical basis for constructs such as the holographic principle *[16, 17]*.

*5.2 Shortcomings of the Model*
The graviton radiation model has structural shortcomings and assumptions that should be described, if only to put the variation in graviton wavelengths calculated for the different cases considered into perspective. The most significant is that there is currently no physical reason to suspect that photons will ever fuse together at such high energies. This does not necessarily imply that the assumptions used to describe the graviton scattering mechanism are wrong; the Standard Model has not been extended, tested, or verified at these energy levels. Therefore, the conjecture that photons can fuse at these energy levels into stable, four-dimensional, di-photonic structures may now be considered a derived prediction (predicated on the successful corroboration of the graviton scattering mechanism employed in this model with expectations derived from Big Bang theory).

The remaining shortcomings of the model relate to the mechanics of the graviton scattering model itself. The characteristic absorption length of an object, dx, is not the same for all incidence angles. Non-homogeneous graviton emitting structures (such as the Earth) would also influence the net characteristic absorption length, and this is a second source of uncertainty. For an object such as the Earth, the densest regions of the Earth's oblate spherical shape are located deep within the Earth. The model also makes no distinction as to the inertial reference frames of the different massive bodies, and so it has no integration with the Special or General Theories of Relativity. However, there is nothing to preclude this integration identified thus far.

A final shortcoming of this model is that it is a very simple model of graviton-graviton interaction. There is no assumed variation in graviton cross-section assumed for different particles, or for particles in different reference frames. The calculated graviton cross-section is based on bulk properties of its constituent photons derived from Wien's Displacement Law, thus the scalar relationship of cross-section with the square of the wavelength described in this paper may not accurately reflect the actual cross-section of a graviton (though, as a first approximation, it is satisfactory for the calculations made in this paper to confirm the utility of the model).

## 6. CONCLUSIONS

The graviton scattering model described in this paper functions by relating two different fundamental properties of the graviton in such a way as to yield an estimate of the average wavelength of the graviton. For the Earth-Human, Moon-Human, and Earth-Moon systems considered in this paper, all results were essentially approximate to the Planck length, even with all of the uncertainties associated with the construction of the models. The basic framework employed through the use of this model suggests the tantalizing possibility that physically comprehensible systems of graviton-graviton interaction may underlie the mechanism that gives rise to gravity, and perhaps Heisenberg's Uncertainty Principle.

Predictions made on the basis of the calculated results of this model are that the isotropic background density of gravitons in space is very high, that the quantum properties of gravitons give rise to a minimum cutoff size of gravitationally-dominated solid bodies, and that for distances roughly equal to or shorter than the range of the strong nuclear force the graviton background produces quantum effects.